\documentclass{article}
\usepackage[utf8]{inputenc}
\usepackage{graphicx}
\usepackage{color}
\usepackage{subfigure}
\usepackage[
backend=bibtex,
style=numeric,
citestyle=numeric
]{biblatex}
 
\usepackage{geometry}
 \usepackage{authblk}

\providecommand{\keywords}[1]{\textbf{\textit{Keywords: }} #1}

\title{Smart Grid Co-Simulation with MOSAIK and HLA: A Comparison Study}
\author[1]{C. Steinbrink}
\author[2]{A. A. van der Meer}
\author[2]{M. Cvetkovic}
\author[1]{D. Babazadeh}
\author[3]{S. Rohjans}
\author[2]{P. Palensky}
\author[1]{S. Lehnhoff}
\affil[1]{OFFIS -- Institute for Information Technology \\
              Oldenburg, Germany\\
              [steinbrink, babazadeh, lehnhoff]@offis.de}
\affil[2]{Delft University of Technology\\
              Delft, The Netherlands\\
              [a.a.vandermeer, m.cvetkovic, p.palensky]@tudelft.nl}
\affil[3]{Hamburg University of Applied Sciences \\
              Hamburg, Germany\\
              sebastian.rohjans@haw-hamburg.de}
\date{October 2017}

\begin{document}

\maketitle

\begin{abstract}
Evaluating new technological developments for energy systems is becoming more and more complex. The overall application environment is a continuously growing and interconnected cyber-physical system so that analytical assessment is practically impossible to realize. Consequently, new solutions must be evaluated in simulation studies. Due to the interdisciplinarity of the simulation scenarios, various heterogeneous tools must be connected. This approach is known as co-simulation. During the last years, different approaches have been developed or adapted for applications in energy systems. In this paper, two co-simulation approaches are compared that follow generic, versatile concepts. The tool \textsc{mosaik}, which has been explicitly developed for the purpose of co-simulation in complex energy systems, is compared to the High Level Architecture (HLA), which possesses a domain-independent scope but is often employed in the energy domain. The comparison is twofold, considering the tools' conceptual architectures as well as results from the simulation of representative test cases.
It suggests that \textsc{mosaik} may be the better choice for entry-level, prototypical co-simulation while HLA is more suited for complex and extensive studies.

\keywords{Co-simulation, \textsc{mosaik}, HLA, Cyber-physical energy systems}
\end{abstract}

\section{Introduction}
\label{sec:intro}
The evaluation of new planning, operation and control designs in the energy domain requires thorough analysis.
In the recent years, the overall character of the domain has turned to the field of cyber-physical energy systems (CPES) with smart grids as one of its most prominent concepts.
This development leads to complex, interdisciplinary setups that are infeasible to handle with purely analytical evaluation.
Since testing new approaches in laboratories or in the field is too inflexible and expensive for early development stages, it is an established procedure to conduct co-simulation beforehand.
This technique is defined as the coordinated execution of two or more simulation models that differ in their representation as well as in their runtime environment \cite{schloegl2015}. 

The first co-simulation implementations have mostly been focused on analysis of the interaction between power system and communication network simulation models \cite{lin2011,godfrey2010,georg2013,mets2011}. 
However, besides this direct coupling of two tools, more generic co-simulation approaches have been developed during the last years. 
They are called \emph{co-simulation frameworks} because one of their integral parts is a middleware that is responsible for data exchange and temporal synchronization of several simulation models.
Many prominent examples for such frameworks are either based on the \textsc{Ptolemy II} software \cite{ptolemaeus2014} (e.\,g. the Building Controls Virtual Test Bed \cite{Wetter2011}) or on various implementations of the High-Level Architecture standard (HLA) for distributed simulation systems \cite{dahmann1997} (e.\,g. C2WT-TE \cite{neema2016}).
A rather new example for a CPES co-simulation framework is the tool \textsc{mosaik} that has been developed at the Oldenburg Institute for Information Technology, OFFIS \cite{schutte2011,rohjans2013}.
In contrast to Ptolemy- or HLA-based systems, it features a more concise set of functionalities, and is aimed at being easy to apply for users from different domains.

Due to the complex character of co-simulation frameworks, it can be difficult for CPES researchers to assess which tool suits their needs.
This paper is aimed at facilitating this task via a thorough comparison of \textsc{mosaik}, as a light-weight tool on the one hand, and an implementation of HLA, as a representation of the popular standard on the other hand.
This comparison is based on both, the particular architectural concepts of the tools as well as results from a simulation study implementing a representative CPES setup. 

The remainder of this paper is structured as follows: 
Section~\ref{sec:cosim} features a discussion of co-simulation in CPES in general as well as \textsc{mosaik} and HLA in particular. 
The conceptual tool comparison is conducted in Section~\ref{sec:conceptcomp}, followed by the co-simulation study described in Section~\ref{sec:cosimstudy}. 
Section~\ref{sec:conclusion}, finally, concludes the paper.

\section{Co-Simulation in Cyber Physical Energy Systems}
\label{sec:cosim}
Co-simulation in CPES is a complex topic that involves various aspects.
A comprehensive overview of the fundamental concepts is provided in \cite{Palensky2017}.
Some of the most basic concepts are summarized in the following.

A co-simulation setup typically consists of independently executable \emph{simulators} that represent different components/domains of the simulated system.
Furthermore, interfaces are implemented to connect the simulators to the framework, as well as a middleware that organizes the communication between the simulators.
This middleware can either include a so-called \emph{master algorithm} that organizes the complete co-simulation process -- or only a set of communication services so that the co-simulation process emerges from the interaction of the simulators.
The temporal synchronization of simulators may either be realized in an iterative manner or non-iteratively (explicitly).
Since CPES co-simulation often involves closed source and other black box simulators, explicit coupling schemes are usually more widely applied.
In these schemes, simulators are either handled one after the other (\emph{serial}) or in \emph{parallel}.

Since CPES co-simulation involves various different research domains, the demand for standardization is high, especially for the interfacing of simulators.
A popular standard is the \emph{Functional Mock-up Interface} (FMI) \cite{Blochwitz2009} that allows embedding of simulators in so-called Functional Mock-up Units (FMU).
Due to the standard's popularity, it is supported by several tools as well as specifically designed toolboxes (e.\,g. \cite{Widle2013}).

\textsc{Mosaik} and HLA, the tools discussed in this paper, both follow the presented understanding of co-simulation.
Their specific designs are reviewed in the following.

\subsection{mosaik}
\label{sec:mosaik}
The \textsc{mosaik} framework has been designed specifically for CPES/smart grid research with a special focus on the flexible creation of large-scale system configurations that may serve for testing of control strategies \cite{Schutte2013}.

The architecture of \textsc{mosaik} consists of a simulator management module (\emph{sim-manager}) and a \emph{scheduler}.
The sim-manager enables data exchange with simulators by establishing TCP connections with them.
The scheduler, on the other hand, coordinates the exchange of data between all connected simulators based on a common simulation clock. 
In its current state, the scheduler provides discretely-timed, explicit simulator coupling.

The \textsc{mosaik} system is completed by two APIs for user interaction: the \emph{Component-API} and the \emph{Scenario-API}.
The Component-API has to be implemented for each simulator that is connected to \textsc{mosaik}.
It sets up a TCP socket and organizes the data exchange with \textsc{mosaik} in the JSON format.
Various versions of the API are available for different programming languages like Java, Python, or MATLAB.
This way, model developers may implement the interface in the language that is most suitable for their simulator.
The implementation itself is conducted by providing a \emph{meta description} for the simulator: a high-level description of the provided simulation models as well as their variables.
Furthermore, a small number of interface functions have to be implemented that are used by \textsc{mosaik} to control the simulator.
Note that non-simulator components like database or data analysis systems may be integrated into \textsc{mosaik} using the same API.
A mapping between the Component-API and FMI also allows the integration of FMUs into \textsc{mosaik} co-simulation \cite{Rohjans2014}, \cite{VanderMeer2017}.

The Scenario-API, finally, provides a set of commands that allows users to instantiate model entities from the integrated simulators and establish connections between them.
The \textsc{mosaik} user is to employ these functions in a \emph{scenario} script that may then be executed to run the co-simulation process.

\subsection{High Level Architecture}
\label{sec:HLA}
HLA is a general-purpose architecture for distributed simulation. It has been developed in the mid-nineties following an initiative of the United States Department of Defense for the purpose of combat simulation coupling~\cite{Dahmann1998}. The first standardization of HLA came in the year 2000 when IEEE declared it as the IEEE standard 1516 for modeling and simulation~\cite{HLA2000b}. Since then, the HLA capabilities have been expanded.
Its current version can be found under the IEEE standard 1516.2-2010~\cite{HLA2010b}. 

HLA has been designed for coupling of highly autonomous simulators while giving ample control to the user. Using the HLA terminology, these simulators are referred to as \emph{federates} while the entire co-simulation setup is referred to as a \emph{federation}.
The communication between federates is realized via a message-bus, the so-called \emph{Runtime Infrastructure} (RTI).
Through seven groups of services, HLA provides synchronization of federates and message passing via the RTI.
For description of the exchanged data, HLA demands the specification of \emph{Simulation Object Models} (SOM) on the federate level and a \emph{Federation Object Model} (FOM) on the federation level.
These structures are defined using the \emph{Object Model Template} (OMT).
Finally, HLA specifies a set of ten design rules for the creation of federations (rules 1 to 5) and federates (rules 6 to 10).
The synchronization of the entire federation depends on the combination of the particular synchronization mechanisms provided by its federates. Thus, various synchronization mechanisms can be implemented with the proper invocation of HLA services, some of which will be tested later in this paper. 

HLA is capable of synchronizing time-stepped and discrete event simulators. 
The first type of simulators are not event-responsive. 
When a time-stepped simulator reaches a synchronization point, it updates its coupling variables based on the latest message that is intended for it (while any message in between synchronization points except the latest one is not received). 
The second type of simulator, the event-responsive simulators, is interrupted in its time-progression if there is a message intended for it. 
Using the corresponding synchronization mechanism, an event-responsive simulator receives all messages that come its way. 
With HLA, the simulators can choose one or the other type of operation with every step forward that they take.

Finally, the publisher-subscriber mechanism of HLA results in loose coupling between simulators, which allows simple topological reconfiguration of the simulated system during runtime. 

\section{Conceptual Tool Comparison}
\label{sec:conceptcomp}
In terms of their conceptual architectures, HLA and \textsc{mosaik} display some structural analogies. 
In both setups, the simulated system is divided into subsystems that are represented by federates (HLA) or simulators (\textsc{mosaik}), respectively. 
The data exchange between the subsystems is managed by a central instance: the RTI in HLA on the one hand, and \textsc{mosaik}’s software core on the other hand, consisting of the sim-manager and the scheduler. 
Furthermore, the communication between the RTI and the federates is handled via a number of standardized function calls, similar to the connection between \textsc{mosaik} and its simulators.

Aside from these architectural similarities, however, HLA and \textsc{mosaik} imply workflows of different nature for their users. 
There are typically two types of tasks a user may assume when working with a co-simulation environment. 
The first one is the integration of a new component into the environment. 
The second one is the creation of a co-simulation study employing the already integrated components. 
Both of these tasks require different specification and implementation work in HLA and \textsc{mosaik}.

Integration of new simulators into \textsc{mosaik} requires, as mentioned before, implementation of the Component-API. 
This involves the specification of the meta description and the implementation of interface functions. 
For integration of new federates into HLA, on the other hand, a SOM has to be specified and a so-called \emph{federate ambassador} implemented that employs RTI services.

\textsc{Mosaik}'s meta description is a very high-level simulator representation that simply specifies the simulation models that may be instantiated from the simulator, their parameters that may be adjusted by the user, and their attributes that may serve for data exchange with other simulators. 
No specification of units, data types or variability is given for the attributes (in contrast to the FMI standard). 
Instead, users are expected to care for attribute compliance themselves when coupling simulators. 
The SOM of HLA, in contrast, is much more extensive. 
Similar to the meta description, it includes information about objects and attributes provided by the federate. 
However, the SOM descriptions are much more detailed, incorporating information about data types, units, resolution, accuracy, and so forth.

For the interaction between \textsc{mosaik} and a simulator, four interface functions are needed that have to be implemented by the user. 
They are then called by the \textsc{mosaik} software core to assume the basic tasks of 1) initializing a simulator, 2) instantiating and parameterizing simulation models, 3) providing input and advancing the simulator in time, 4) and requesting the simulator's output data. 
In contrast, HLA provides a much wider range of functions for the coordination between federates and the RTI which allows for more nuanced interfacing.
The functions are not implemented by the user but provided by the RTI as services.
The more than 20 services are grouped into clusters like federation management, object management or time management. 
In order to employ the desired services, the user has to implement the already mentioned federation ambassador that sends calls to the RTI and receives callbacks from it.
Another major purpose of the discussed systems is the create of executable co-simulation setups.
For the creation of a co-simulation scenario in \textsc{mosaik}, the user has to write a scenario script in Python, employing commands from the Scenario-API. 
This script should specify which integrated simulators are used, how many model entities are instantiated, and how they are interconnected.
Furthermore, data values have to be given for the parameterization of simulators and model entities, as well as the length of the simulated time period. 
The interconnections may be guided by rules that allow filtering by entity types or previously made connections. 
This way, even complex connections between large model entity sets can be established with only a few lines of code.

Co-simulation experiments in HLA are less centrally defined than in \textsc{mosaik}. 
As indicated before, HLA federates are more autonomous than \textsc{mosaik} simulators, and thus, retain more control over the simulation process. 
For example, they possess functionalities to dynamically connect to or disconnect from a running simulation. 
This way, a co-simulation experiment is strongly defined by the configuration of the individual federates. 
Nevertheless, the HLA specifies a federation also via some central structures. 
The most important structure for setting up a federation is a FOM. 
Like the SOM, it is based on the OMT, but while the SOM describes the attributes and objects of a single federate, the FOM describes all attributes and objects that may be used for data exchange within the federation. 
SOMs are in some sense subordinate to the FOM. 
In general, a FOM is designed for a specific application domain and can be reused for new simulations in the same domain. 
Aside from a FOM, federations may be complemented by so-called \textit{federation agreements} or scenarios definitions for documentation purposes, but these concepts are not specified by HLA rules or official templates. 

\begin{figure}[t]
  \centering
  \includegraphics[width=0.4\textwidth]{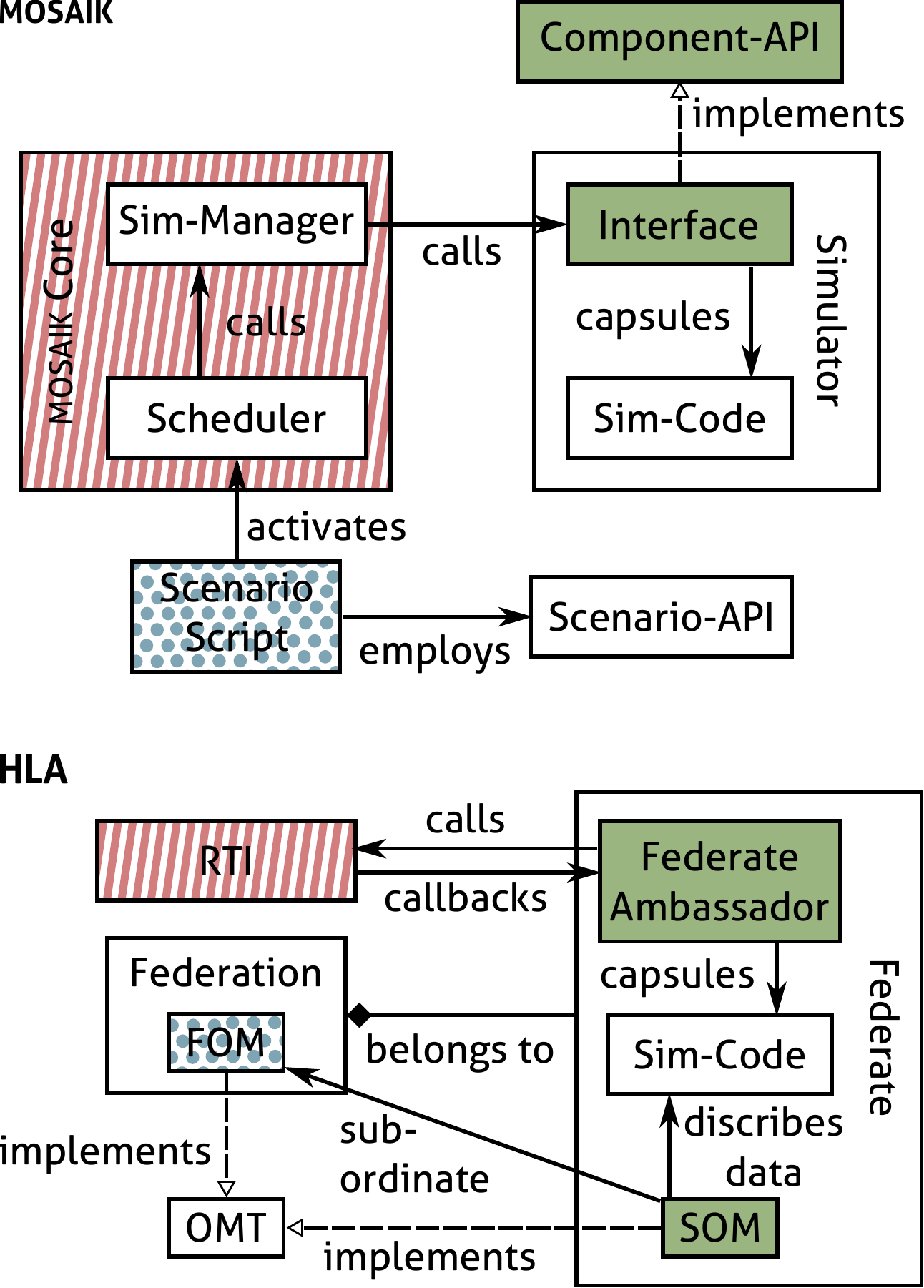}
\caption{Architecture overviews of \textsc{mosaik} and HLA.}
\label{fig:archs}       
\end{figure}

All in all, \textsc{mosaik} and HLA both follow the common structure of co-simulation systems possessing individual simulator components that communicate via a message bus.
This is reflected in the schematic architectural overview in Figure~\ref{fig:archs}.
Similarities are given in the form of a communication hub (red, striped), component interfacing and description (green, solid), and some specification of component interaction (blue, dotted).
Nevertheless, the concepts also differ in some key design principles. 
As indicated above, a HLA federate is potentially more versatile than a \textsc{mosaik} simulator and can intervene with the simulation flow more autonomously. 
Accordingly, the course of the simulation execution is more loosely defined in HLA than in \textsc{mosaik} in order not to limit the capabilities of the federates. 
While \textsc{mosaik} demands a scenario script that specifies fixed interconnections between simulators, HLA merely defines common data structures via the FOM. 
All other aspects of the federation execution are defined by the federate behaviors and publication-subscription scheme for data exchange.

The main differences and similarities between \textsc{mosaik} and HLA are also reflected in the ten design rules for federates and federations that are part of the HLA standard. 
The first five rules represent requirements for the design of federations. 
Their notions are largely mirrored by \textsc{mosaik}. The rules 2 to 4 especially illustrate the architecture of the HLA. 
They postulate:
\begin{itemize}
\item All objects in the federation shall belong to federates and not to the RTI.
\item Data exchange during federation execution shall happen exclusively between federates and the RTI.
\item Communication between federates and the RTI shall only be executed via the services defined in the standard.
\end{itemize}
All of these principles are also true for \textsc{mosaik} when replacing the terms "federate" and "RTI" with "simulator" and "sim-manager", respectively. 
This underlines the basic architectural similarity between the two concepts. 
In contrast, the HLA rules 6 to 10 specify the federate design and illustrate some of the major differences between \textsc{mosaik} and HLA.
These rules grant a lot of autonomy to federates. 
For example, they shall be able to update their conditions for data exchange or transfer ownership of objects between each other. 
Such concepts are not stipulated in \textsc{mosaik}. 
Furthermore, federates have more power over their time management in HLA. 
Instead of being externally stepped like in \textsc{mosaik}, they may request time advancement at the RTI. 
Depending on the type of request, timing may be negotiated between federate and RTI, allowing for more elaborate options for temporal coupling.

In summary, the conceptual comparison reveals that HLA provides a more versatile co-simulation framework than \textsc{mosaik}. 
Its federates may possess a wider variety of capabilities than \textsc{mosaik} simulators so that more different forms of interaction are possible. 
This is especially true for time synchronization. 
Since HLA allows for negotiation of time advancement, iterative coupling is possible, which so far is not provided by \textsc{mosaik}. 
However, it has to be noted that the versatility of HLA comes at the cost of a more complex implementation process. 
Since \textsc{mosaik}'s Component-API is more concise than its HLA equivalent, its usage is easier to learn.
Furthermore, creating an executable co-simulation study in \textsc{mosaik} requires only one script with the help of the Scenario-API. 
The corresponding process in HLA is more loosely defined and typically requires individual configuration of the federates. 
Therefore, deciding between the usage of \textsc{mosaik} and an implementation of HLA entails a trade-off between a flat learning curve (\textsc{mosaik}) and a wider field of application (HLA). 
In addition, it has to be considered whether simulators are to be treated as black boxes or should be freely configurable. 
The former point follows the \textsc{mosaik} philosophy while the latter stands of the HLA design.
The conceptual comparison of the two systems is summarized in Table~\ref{tab:compare}.
\begin{table*}
\caption{Overview of comparison between \textsc{mosaik} and HLA.}
\label{tab:compare}       
\centering
\begin{tabular}{lll}
\hline\noalign{\smallskip}
Category & \textsc{mosaik} & HLA  \\
\noalign{\smallskip}\hline\noalign{\smallskip}
\shortstack[l]{Communication \\ \hspace{1mm}} & \shortstack[l]{Handled by Sim-Manager; \\ simulators are called} & \shortstack[l]{Handled by RTI; federates send requests \\ and receive callbacks}\\
\vspace*{0.2mm}\\
\shortstack[l]{Components \\ \hspace{1mm}} & \shortstack[l]{Simulators with limited capabilities; \\ user implements interface functions} & \shortstack[l]{Very flexible federates; \\ user employs RTI services} \\
\vspace*{0.2mm}\\
\shortstack[l]{Time \\ Synchronization} & \shortstack[l]{Organized by scheduler; \\ discrete time, variable step size} & \shortstack[l]{Individually defined for the federates \\ \hspace{1mm}} \\
\vspace*{0.2mm}\\
Data Exchange & Users take care of exchange validity & Validity defined by SOMs and FOM \\
\vspace*{0.2mm}\\
\shortstack[l]{Simulation \\ Configuration} & \shortstack[l]{Defined in executable script \\ via Scenario-API} & \shortstack[l]{Defined via individual federate \\ message subscriptions} \\
\noalign{\smallskip}\hline
\end{tabular}
\end{table*}

\section{Co-Simulation Study}
\label{sec:cosimstudy}
It has been stated above that the HLA design allows for more types of simulator coupling than \textsc{mosaik}. 
Therefore, comparing the two systems in the context of a co-simulation study can be misleading when being based on incomparable synchronization schemes. 
However, if analogous schemes are selected, results produced with \textsc{mosaik} and HLA should be equivalent to each other. 
Studying this hypothesis is the purpose of this section. 
Two simulators have been coupled using only those schemes that are applicable in HLA as well as in \textsc{mosaik}. 
Such a limitation is reasonable since selection of the coupling scheme is influenced by capabilities and limits of the simulators in question. 
In other words, simulators are assumed here that are not compatible with HLA's more elaborate coupling approaches and thus may just as well be handled by \textsc{mosaik}.

As a HLA implementation the software tool CERTI \cite{certi} has been used, version~3.5.1.
Just as \textsc{mosaik}, it follows an open source licensing model.
The employed \textsc{mosaik} version is 2.3.0.

\subsection{Test System Configuration}
\begin{figure}
\centering
\includegraphics[]{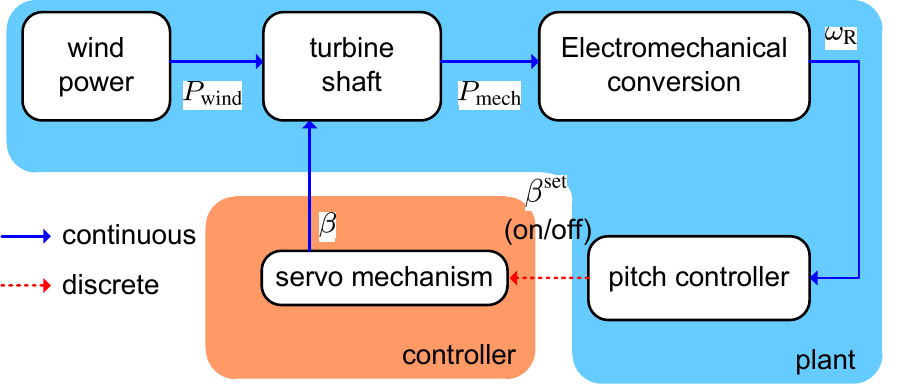}
\caption{\label{fig:WTG-model}Elementary wind turbine model (plant) with active pitch regulation (controller).}
\end{figure}

The capabilities of \textsc{mosaik} and HLA will be compared by simulating the storm control of a wind turbine generator. 
This protection mechanism is engaged when the blade tip speed, and hence rotor frequency violates a threshold value, which commonly equals the rotor speed at rated power output. 
The test system is shown in Fig.~\ref{fig:WTG-model}. 
Although such a small system does not call for advanced modeling or simulation methods, the system contains both continuous and discrete behavior, and thus is considered a good and comprehensible example for the comparison of \textsc{mosaik} and HLA.

It consists of a wind power model simulating a sudden wind gust, a turbine shaft model, a second-order plant model of the electromechanical conversion, a discrete pitch controller, and a servo mechanism for blade pitching. In case the rotor speeds exceeds 110 \% of the rated value, the pitch controller enables the servo mechanism $\beta^\textrm{set}$.  The servo mechanism on its turn gradually increases the pitch angle $\beta$, decreasing the turbine shaft power.  

Engaging the servo mechanism is a sudden event based on the physical behavior of the wind turbine. From a co-simulation viewpoint, it is interesting to see how such an event is processed between different simulators/federates. Therefore, the test system is split into a plant simulator producing the event and a controller simulator, which responds to it, using $\beta^\textrm{set}$ as an interface attribute.  

\subsection{Co-Simulation Setup}
\label{sec:setup}

The plant as well as the controller simulation model have been created in Modelica and exported as FMUs.
A monolithic reference simulation has been conducted using the combination of both models within Modelica.

As mentioned in Section~\ref{sec:cosim}, the basic, non-iterative coupling between two simulators may either be serial or parallel.
These are the only options for simulator coupling supported by \textsc{mosaik} whereas HLA may also allow extensions like stepping negotiation or \emph{look-ahead} calculations (if they are supported by the simulators in question). 
A look-ahead can be implemented in HLA by making the RTI time lag behind the simulator time by one step.
More explanation on this topic will follow further below.

Aside from the general synchronization mechanism, the temporal resolution (i.e. the time step size) of the individual simulators plays a crucial role in their coupling. 
Both, \textsc{mosaik} and HLA, allow simulators to employ adaptive time step-sizes.
For the sake of comparability, however this example study utilizes a fixed time step-size. Two basic cases of discretely timed co-simulation are examined: 1) both simulators have the same step size, and 2) the simulators have different step sizes. 
Combining these setups with the two synchronization options, six test cases (TC) have been established and realized with both, \textsc{mosaik} as well as HLA (see Tab.~\ref{tab:cases}).
In the serial setup (TC\,4-6), the plant simulator is always stepped before the controller.
\begin{table}
\caption{Test cases for system comparison.}
\label{tab:cases}       
\centering
\begin{tabular}{l|cc}
\hline\noalign{\smallskip}
 & \shortstack{Parallel \\ Setup} & \shortstack{Serial \\ Setup}\\
\noalign{\smallskip}\hline\noalign{\smallskip}
Both step sizes: $0.02\,s$ & TC\,1 & TC\,4\\
\vspace*{0.2mm}\\
Both step sizes: $0.015\,s$ & TC\,2 & TC\,5\\
\vspace*{0.2mm}\\
\shortstack[l]{Plant step size: $0.02\,s$, \\ Controller step size: $0.015\,s$} & \shortstack{TC\,3 \\ \hspace{1mm}} & \shortstack{TC\,6 \\ \hspace{1mm}}\\
\noalign{\smallskip}\hline
\end{tabular}
\end{table}

\subsection{Results}
\label{sec:res}
\begin{figure}[t]
\centering
  \includegraphics[width=0.6\textwidth]{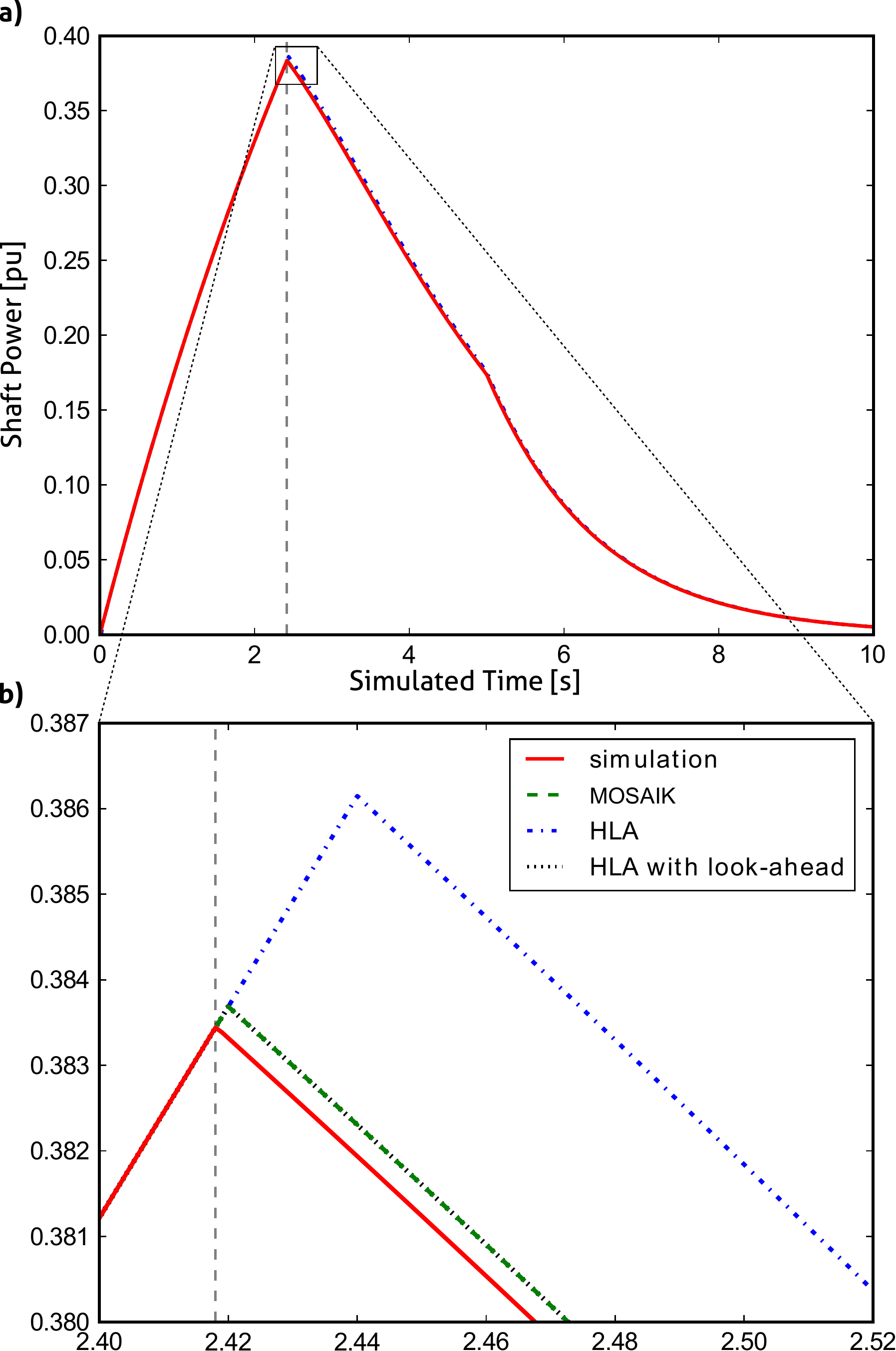}
\caption{Calculated wind power input to the turbine for different (co-)simulation setups. Full plot (a) and zoomed-in around the time of the event (b). Co-simulation is conducted with the setup of TC4.}
\label{fig:blades}
\end{figure}

The behavior of the simulated system is characterized by the wind power (and hence shaft speed) exceeding the specified threshold which leads to a response of the control mechanism.
The shaft power (in p.u. of the actual wind power) reaching the turbine is plotted over time in Figure~\ref{fig:blades}\,a.
It can be seen that the turbine input increases quickly during the first two simulated seconds until the threshold is reached (gray, dashed line).
After that, the interference of the controller leads to a decrease of the wind power reaching the turbine.

\begin{figure*}[t]
  \includegraphics[width=\textwidth]{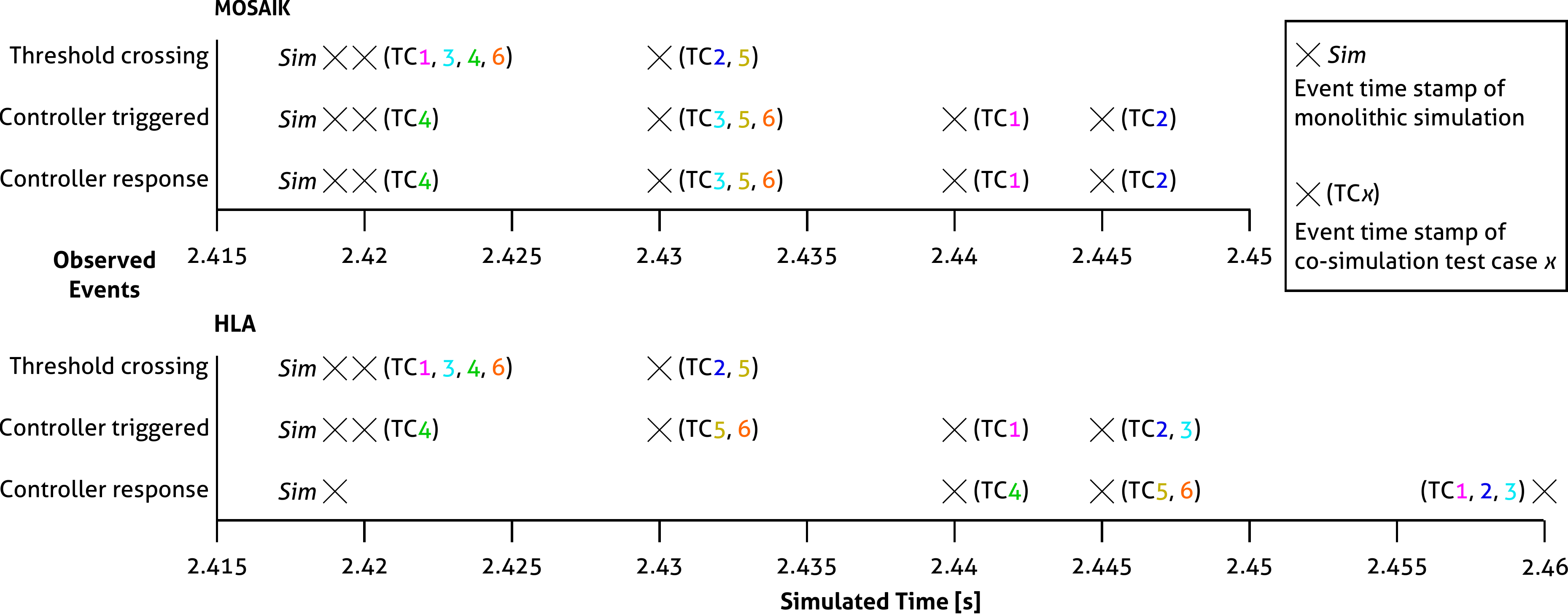}
\caption{Timing of events in the conducted co-simulation test cases.}
\label{fig:timing}
\end{figure*}

Next to the threshold crossing, two more events can be defined, related to this system behavior.
For one, the information of the threshold crossing reaches the controller to trigger it.
The final event is the response of the controller.
In the monolithic simulation with a temporal resolution of $0.001\,s$, these three events are processed within the same time step.
In the co-simulation setups, in contrast, this is not necessarily possible since the system is split between the plant and the controller.
In fact, the conducted co-simulation experiments reveal that the timing of the events depends on the employed synchronization scheme in combination with the temporal resolution of the simulators.
Figure~\ref{fig:timing} displays the time stamp of the three discussed events for every conducted test case.
Each time stamp is indicated by an $\times$.
It can be seen that the time stamps produced by the monolithic simulation (Sim) are $2.419$ for all three events.
Since this is no multiple of the resolutions of the separate simulators, none of the co-simulation test cases could reproduce these time steps.
Even more so, the events cannot be processed within the same time step for some test cases due to the data exchange between the simulators.

A number of interesting observations can be made when comparing the time stamps of the events for the different test cases.
First of all, the serial synchronization setup with a time step of $0.02\,s$ (TC4) produces the best output of all the analyzed co-simulation setups.
This can be explained by the order of the events.
Since the plant simulator registers the first event and then triggers the controller, all events can be processed within the same time step if the co-simulation setup features a serial scheme that advances the plant before the controller.
Interestingly, test cases with a time step size of $0.015\,s$ tend to reproduce the timing of the events less accurately despite the higher resolution.
This is explained by the fact that the actual event times are closer to a multiple of $0.02\,s$ than of $0.015\,s$.
In other words, due to the system's dynamics, a higher temporal resolution does not necessarily lead to more accurate results -- if it is not at least a magnitude higher.
Another interesting observation is the fact that all test cases conducted with \textsc{mosaik} reproduce the event times slightly better than those conducted with HLA.
The explanation for this lies in the type of HLA services used for this comparison. 
So-called \emph{non-zero look-ahead} services have been used. 
These services stipulate that a non-zero look-ahead time must be provided by the federate to the RTI in order to step forward in time. 
Under such rules, the message sent by one simulator to another will take non-zero time to be delivered to its destination. 
Thus, while mosaik allows for a message to be sent and delivered at the same time instance, the particular set of HLA services used for this comparison adds a delay of one time step between the time the message is sent and the time it is delivered.

The discrepancy between the \textsc{mosaik} and the HLA results can also be seen in Figure~\ref{fig:blades}\,b (zoom).
For both co-simulation systems, the results of TC4 are compared with those of the monolithic simulation (red, continuous line).
While none of the co-simulation systems is able to reproduce the simulation results perfectly, the \textsc{mosaik} results (green, dashed line) approximate them more closely than the HLA results (blue, dash-dotted line).
However, this is only true for a HLA setup that displays similar limitations as \textsc{mosaik}.
As mentioned before, HLA federates may be equipped with more capabilities than \textsc{mosaik} simulators.
Introducing the ability to conduct look-ahead operations into the federate design leads to a HLA setup that matches the output of the \textsc{mosaik} co-simulation (black, dotted line, that is overlaid on top of the green, dashed line).
After all, the look-ahead is able compensate the non-zero look-ahead aspect of the HLA services.
Furthermore, it has to be noted that newer versions of HLA include \emph{zero look-ahead} services and thus should be able to match the \textsc{mosaik} results without the need to implement look-ahead operations.

\section{Conclusion}
\label{sec:conclusion}
The presented work gives a concise overview of structural similarities and differences between the HLA and the \textsc{mosaik} concept in the context of co-simulation. 
CPES researchers that seek to employ co-simulation may use the work as a guideline to decide which tools suits their needs. 
One major difference between \textsc{mosaik} and HLA lies in the interfacing of simulators (or federates, respectively). 
HLA federates are generally more autonomous and powerful than \textsc{mosaik} simulators which allows for more versatile coupling, but at the same time leads to a higher interfacing effort. 
Next to this aspect, the two systems differ in the way they specify interaction between simulators/federates. 
In HLA, users define federations that may involve complex interactions but are time-consuming to set up, requiring a FOM to specify all valid interaction types and exchanged data, and user developed scripts to invoke the RTI services in an appropriate order. 
\textsc{Mosaik}, on the other hand, requires users to define scenarios that only allow for basic data exchange. 
However, they are easily established and allow large-scale topology changes with a set of simple commands.

The conducted co-simulation experiments reveal that a setup with an older HLA version may produce less accurate results than \textsc{mosaik} if both systems employ only basic synchronization schemes.
The difference results from non-zero look-ahead services being employed in the HLA setup.
However, this discrepancy may be reconciled by implementing a look-ahead in the federates or using a current version of HLA that provides zero look-ahead services.
The general accuracy that may be achieved with co-simulation depends strongly on the behavior of the simulated system in combination with the employed synchronization scheme and the temporal resolution of the simulators.

Both framework instances used in this study follow an open source licensing model and are thus freely available to researchers and possibly even customizable.
As indicated before, a variety of HLA implementations exist, some of which are commercial solutions and might thus come with a greater user comfort.

In summary, \textsc{mosaik} allows for an easier entry into co-simulation while HLA is more versatile and powerful on the long run. 
Therefore, researchers may want to employ \textsc{mosaik} in early stages of a co-simulation study when large numbers of different simulators and topologies need to be tested. 
HLA, on the other hand, may be applied for more mature studies that require extensive control over simulator coupling for the sake of output accuracy. 
In future work, a modular approach for simulator interfacing may be developed that supports the transition between the two styles of co-simulation by allowing gradual extension of the same interface.
Since standardization is an important aspect for acceptance and usability of such an approach, FMI should be utilized further for interfacing.
Next steps in the work with HLA and \textsc{mosaik} will involve work with more sophisticated simulators as well as comparison of scalability in terms of system size and temporal resolution.
For some versions of both framework concepts, studies have already been conducted that demonstrate the capability to host simulations with several thousand entities \cite{sonnenschein2012,brito2013}.
However, the associated findings are not necessarily up to date and can hardly be compared due to differences in the setup and employed simulators.
Palminitier and colleagues have established a simple scalability test for the CERTI HLA implementation \cite{Palmintier2017}.
They evaluate the scaling of its performance in regard to problem size and computing resources.
An analogous evaluation setup can be established for \textsc{mosaik}.
Implementing such a benchmarking framework and deriving a comparative performance analysis of the discussed co-simulation tools is part of future work.

\section*{Acknowledgements}
This work is supported by the European Community’s Horizon 2020 Program (H2020/2014-2020) under project ``ERIGrid'' (Grant Agreement No. 654113).

\printbibliography

\end{document}